\documentclass[prl,aps,twocolumn,superscriptaddress]{revtex4}
\usepackage{psfrag,slashed,cancel,array,graphicx,todonotes,hyperref}
\usepackage[utf8]{inputenc}
\usepackage{mathtools}
\usepackage{mathrsfs}
\usepackage{multirow}
\usepackage{braket}
\usepackage{slashed}
\usepackage{booktabs} 
\usepackage[normalem]{ulem}
\usepackage{slashed}
\usepackage{diagbox}
\hypersetup{pdftitle={},pdfcreator={},linkcolor=[rgb]{0.15,0.35,0.75},colorlinks=true,citecolor=[rgb]{0.675,0,0.2},urlcolor=[rgb]{0.15,0.35,0.65}}

\def\beq{\begin{equation}}
\def\eeq{\end{equation}}
\def\bsp#1\esp{\begin{split}#1\end{split}}
\def\d{{\rm d}}

\newcommand{\nn}{\nonumber}

\AtBeginDocument{
\heavyrulewidth=.08em
\lightrulewidth=.05em
\cmidrulewidth=.03em
\belowrulesep=.65ex
\belowbottomsep=0pt
\aboverulesep=.4ex
\abovetopsep=0pt
\cmidrulesep=\doublerulesep
\cmidrulekern=.5em
\defaultaddspace=.5em
}

\def\be{\begin{equation}}
\def\ee{\end{equation}}

\DeclareRobustCommand\slout{\bgroup\markoverwith{\color{blue!30}{\rule[0.4ex]{2pt}{0.8pt}}}\ULon}

\begin{document}

\preprint{XXX}

\author{Liang Dong}
\email{liang.dong@sjtu.edu.cn}
\affiliation{State Key Laboratory of Dark Matter Physics, Shanghai Key Laboratory for Particle Physics and Cosmology, Key Laboratory for Particle Astrophysics and Cosmology (MOE),
School of Physics and Astronomy, Shanghai Jiao Tong University, Shanghai 200240, China}
\author{Shen Fang}
\email{sfang23@m.fudan.edu.cn}
\affiliation{Department of Physics, Center for Field Theory and Particle Physics, and Key Laboratory of Nuclear Physics and Ion-beam Application (MOE), Fudan University, Shanghai, 200433, China}
\author{Jun Gao}
\email{jung49@sjtu.edu.cn}
\affiliation{State Key Laboratory of Dark Matter Physics, Shanghai Key Laboratory for Particle Physics and Cosmology, Key Laboratory for Particle Astrophysics and Cosmology (MOE),
School of Physics and Astronomy, Shanghai Jiao Tong University, Shanghai 200240, China}
\author{Hai Tao Li}
\email{haitao.li@sdu.edu.cn}
\affiliation{School of Physics, Shandong University, Jinan, Shandong 250100, China}
\author{Ding Yu Shao}
\email{dyshao@fudan.edu.cn}
\affiliation{Department of Physics, Center for Field Theory and Particle Physics, and Key Laboratory of Nuclear Physics and Ion-beam Application (MOE), Fudan University, Shanghai, 200433, China}
\affiliation{Shanghai Research Center for Theoretical Nuclear Physics, NSFC and Fudan University, Shanghai 200438, China}
\affiliation{Center for High Energy Physics, Peking University, Beijing 100871, China}
\affiliation{Southern Center for Nuclear-Science Theory (SCNT), Institute of Modern Physics, Chinese Academy of Sciences, Huizhou 516000, Guangdong Province, China}
\author{Yu Jiao Zhu}
\email{yzhu@mpp.mpg.de}
\affiliation{Max-Planck-Institut f\"ur Physik, Werner-Heisenberg-Institut, Boltzmannstrasse 8, 85748 Garching, Germany}

\title{NNLO QCD corrections to hadron production in DIS at finite transverse momentum}

\begin{abstract}
We present the \textit{first} calculation of hadron production in deep-inelastic scattering (DIS) at finite transverse momentum to next-to-next-to-leading order (NNLO) in perturbative QCD. 
To overcome the long-standing challenge of infrared divergences in semi-inclusive processes with identified final state hadrons at finite transverse momentum, we implement the recently developed $q_T$-subtraction framework based on the recoil-free jet definition. 
By utilizing the winner-take-all recombination scheme, we achieve a consistent factorization for hadron-jet associated production, enabling the inclusion of $\mathcal{O}(\alpha_s^3)$ corrections.
Our NNLO results generally demonstrate an improved convergence of the perturbative expansion and a reduction in scale uncertainties compared to previous next-to-leading order ones, especially for comparisons to multiplicity data from the ZEUS Collaboration. 
This work provides a high-precision theoretical foundation for the upcoming electron-ion collider era and establishes a new benchmark for the exploration of the nucleon's three-dimensional structure.
\end{abstract}

\maketitle

\paragraph*{Introduction.---}
The semi-inclusive deep inelastic scattering (SIDIS) process, $l(l) + N(P) \to l'(l') + h(P_h) + X$, is fundamental to mapping the three-dimensional nucleon structure. 
While inclusive DIS has historically provided a precise mapping of collinear parton distribution functions (PDFs), the detection of a final state hadron with finite transverse momentum $P_{hT}$ enables the exploration of the transverse momentum dependent (TMD) dynamics of partons and the hadronization process \cite{Collins:1981uk, Collins:1981uw}. 
In particular, the high-$P_{hT}$ region is of great theoretical interest. 
It provides a unique window into the transition between the non-perturbative TMD framework and the perturbative collinear factorization regime, where $P_{hT}$ is generated by hard radiation.
Over the past decades, experimental programs have provided a detailed mapping of $P_{hT}$ distributions across diverse kinematic ranges. 
Early measurements by the H1 \cite{H1:1996muf, H1:1999iry, H1:2004xgw} and ZEUS \cite{ZEUS:1995acw, ZEUS:1999ies} Collaborations at HERA established the initial benchmarks for inclusive hadron production at high $P_{hT}$. 
These were followed by higher-precision studies from HERMES \cite{HERMES:2012uyd}, COMPASS \cite{COMPASS:2013bfs, COMPASS:2017mvk}, and Jefferson Lab (JLab) \cite{JeffersonLabHallA:2016ctn}, which revealed intricate flavor dependencies and spin-momentum correlations within the nucleon.
Despite these experimental advances, theoretical predictions for the transverse momentum spectrum have largely been restricted to next-to-leading order (NLO) accuracy \cite{Kniehl:2004hf, Daleo:2004pn, Wang:2019bvb}, which exhibit significant scale uncertainties and struggle to provide a simultaneous global description of experimental data \cite{Gonzalez-Hernandez:2018ipj}.
This status contrasts sharply with rapid progress for inclusive quantities in DIS: N$^3$LO corrections have been computed for single jet production in DIS \cite{Currie:2018fgr}, as well as next-to-next-to-leading order (NNLO) corrections for dijet production in DIS \cite{Currie:2017tpe}.
Furthermore, NNLO coefficient functions were recently derived for inclusive SIDIS cross sections~\cite{Abele:2021nyo, Goyal:2023zdi, Bonino:2024qbh, Bonino:2024wgg, Goyal:2024tmo, Goyal:2024emo, Bonino:2025tnf, Haug:2025ava, Bonino:2025qta, Zhou:2025lqv, Bonino:2025bqa, Goyal:2025qyu, Gao:2026tnd}. 
The fully differential description of identified hadrons at finite $P_{hT}$ remains a critical frontier, and achieving NNLO precision is essential to stabilize the perturbative expansion and provide robust theory inputs for the upcoming Electron-Ion Collider (EIC) \cite{Accardi:2012qut, AbdulKhalek:2021gbh}. 
However, extending calculations to this order for final states involving identified particles faces significant hurdles due to the complex cancellation of infrared (IR) divergences \cite{Kinoshita:1962ur, Lee:1964is}. 
For the description of inclusive quantities, various strategies have emerged to resolve these singularities, broadly categorized into local subtraction \cite{Gehrmann-DeRidder:2005btv, Czakon:2010td, Currie:2013vh}, projection-to-Born methods \cite{Cacciari:2015jma}, and slicing techniques \cite{Catani:2007vq, Gao:2012ja, Gao:2014eea, Gaunt:2015pea, Boughezal:2015dva, Berger:2016inr}. 
Yet, applying these techniques to semi-inclusive processes has been limited by the difficulty of consistently treating soft and collinear radiation in the presence of identified hadrons and jets, with a few successful applications~\cite{Fu:2024fgj, Czakon:2025yti, Generet:2025bqx, Gao:2026tnd, Bonino:2026dvr}.

In this Letter, we present the \textit{first} NNLO QCD calculation of identified hadron production in DIS at finite $P_{hT}$. 
The core of our theoretical approach is the implementation of a new $q_T$-subtraction framework recently proposed in Refs.~\cite{Fu:2024fgj, Fu:2026nkd} (see also Refs.~\cite{Buonocore:2022mle, Buonocore:2023rdw, Buonocore:2025ucn}), which generalizes the classical transverse momentum subtraction method~\cite{Catani:2007vq} to final states involving both jets and hadrons. 
By utilizing a recoil-free jet axis, specifically the winner-take-all (WTA) recombination scheme~\cite{Salam:WTAUnpublished, Bertolini:2013iqa}, this method allows for the consistent treatment of radiation from the identified hadron and other jets.
By incorporating $\mathcal{O}(\alpha_s^3)$ corrections through this framework, we demonstrate good convergence of theoretical predictions and reduced theoretical uncertainties. 
This result marks a new era of precision for SIDIS phenomenology and provides a rigorous test of the QCD factorization theorems in the high-$P_{hT}$ domain.

\begin{figure}[t!]
    \centering
    \includegraphics[width=0.4\textwidth]{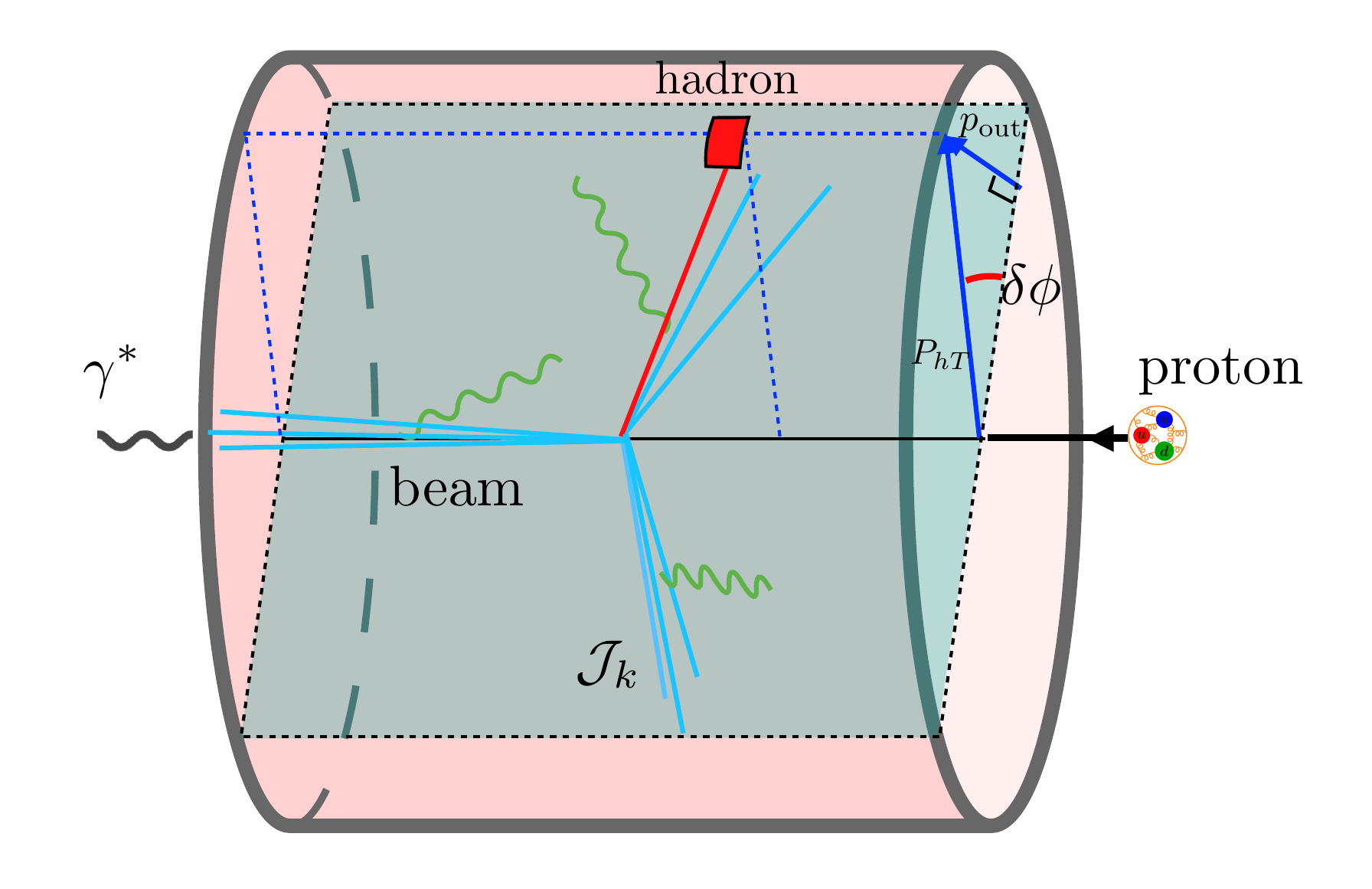}
    \caption{Schematic definition of the kinematic variables used for the $q_T$-subtraction method in the Breit frame. The incoming virtual photon $\gamma^*$ scatters off the proton, producing an identified hadron and a leading recoiling jet ${\cal J}_k$. The blue lines along the proton direction denote beam collinear radiation. The green curved lines indicate the soft radiation.  The slicing variable is defined by the azimuthal decorrelation $\delta\phi$ (or, equivalently, the out-of-plane momentum $p_{\rm out}$ of the hadron with respect to the beam-jet plane). The limit $\delta\phi \to 0$ corresponds to the unresolved back-to-back configuration governed by the factorization formula in Eq.~(\ref{eq:fact_phi_jet}).}
\label{fig:kinematics}
\end{figure}

\paragraph*{$q_T$-subtraction formalism.---}
In this work, we employ the azimuthal decorrelation $\delta\phi$, or, equivalently, the hadron momentum component $p_{\rm out}$ perpendicular to the scattering plane, as the slicing variable. 
The scattering kinematics in the Breit frame are illustrated in Fig.~\ref{fig:kinematics} with the scattering plane defined by the leading recoiling jet and beam directions. 
This partitions the cross section into contributions from an unresolved region, dominated by soft and collinear radiation, and a resolved region characterized by large-angle emissions.
Explicitly, we have
\begin{align}
 {\d\sigma\over \d\mathcal O}= \!\!\int_0^{\delta \phi^{\rm cut}}\d \delta\phi \frac{\d\sigma}{\d\delta \phi\,\d\mathcal O} +\!\!\int_{\delta \phi^{\rm cut}}^{\delta \phi^{\rm max}} \d \delta\phi \frac{\d\sigma}{ \d\delta \phi\,\d\mathcal O}\,,
\end{align}
where $\delta \phi^{\rm cut}$ is a small resolution or slicing parameter introduced in the $q_T$-subtraction method.
The first (unresolved) part is approximated by the leading-power (LP) contribution in the unresolved limit ($\delta\phi^{\rm cut}\to0$), subject to power corrections which vanish in this limit. 
The cancellation of the dependence on $\delta\phi^{\rm cut}$ in the full results serves as a rigorous check of IR safety and numerical stability of our framework.
In the resolved region ($\delta\phi > \delta \phi^{\rm cut}$) where the hadron and jets are well separated, the differential cross sections can be calculated by requiring production of an inclusive hadron in association with a dijet in DIS up to NLO.
We use the one-loop matrix elements for $ei \to ejkl$ and the tree-level matrix elements for $ei \to ejklm$ calculated in Refs.~\cite{Bern:1997sc, Campbell:2002tg, Campbell:2003hd, Campbell:2010ff}, where $ijklm$ represent all possible flavor combinations of QCD partons. 
The jets are reconstructed using the $k_T$ algorithm~\cite{Catani:1993hr, Ellis:1993tq} with the WTA recombination scheme. 
Crucially, the jet algorithm dependence cancels between the resolved and unresolved regions, ensuring the final result is algorithm-independent.
The calculations are carried out in the {\tt FMNLO} framework~\cite{Liu:2023fsq} for identified hadron production, with IR singularities handled using a method of dipole subtraction~\cite{Catani:1996vz} in combination with phase space slicing.

\begin{figure*}[t!]
    \centering
    \includegraphics[width=0.95\textwidth]{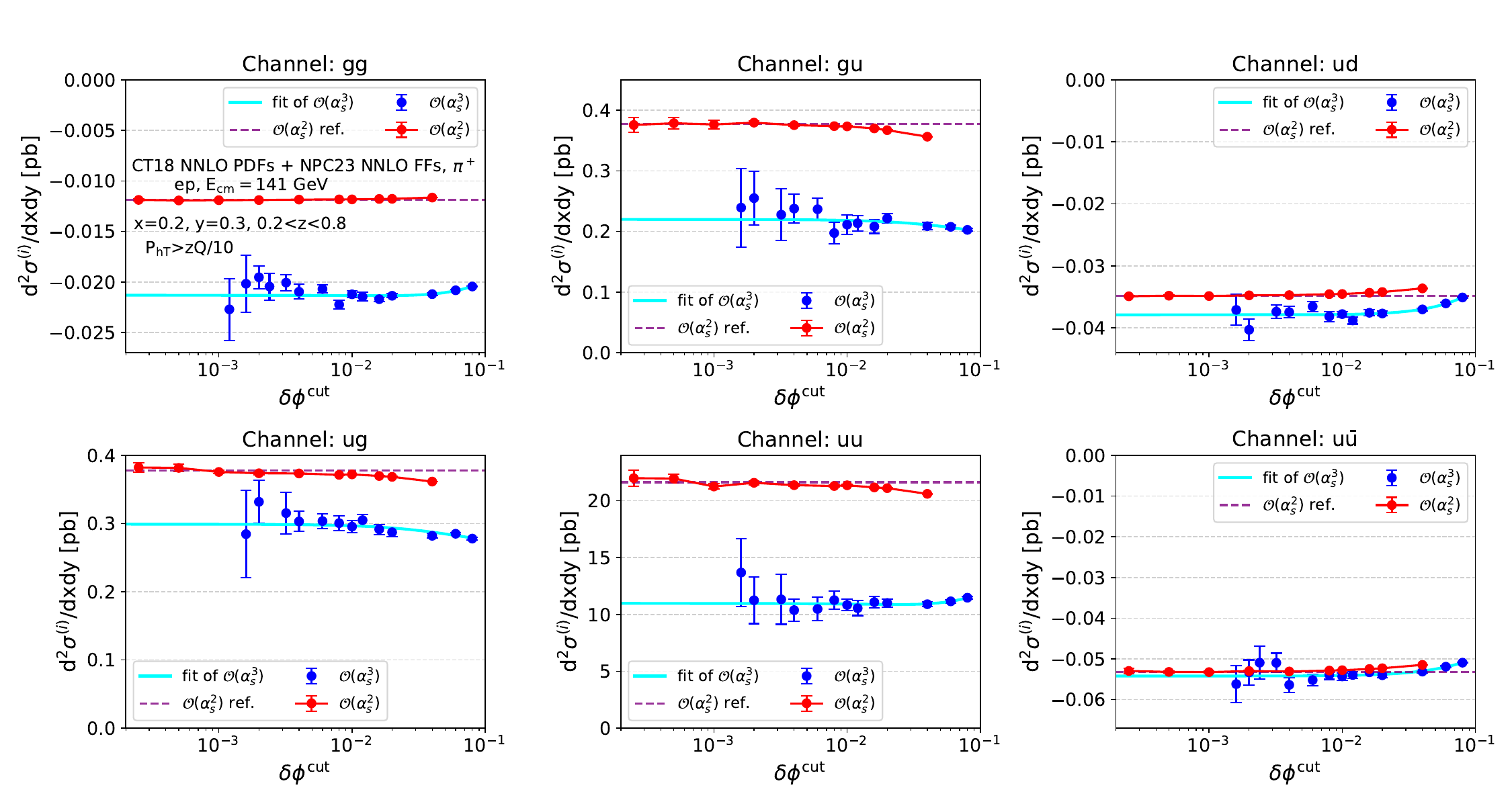}
    \caption{%
    Numerical stability of the total cross section with respect to the slicing parameter $\delta\phi^{\rm cut}$.
    The panels display the $\mathcal{O}(\alpha_s^2)$ (red) and $\mathcal{O}(\alpha_s^3)$ (blue) contributions for six representative partonic channels. 
    Error bars indicate Monte Carlo statistical uncertainties.
    The solid cyan lines represent the fit to the asymptotic limit $\delta\phi^{\rm cut} \to 0$. 
    The dashed purple lines denote the reference values of $\mathcal{O}(\alpha_s^2)$.}
    \label{fig:cutoff_dep}
\end{figure*}

In the unresolved region ($\delta\phi < \delta \phi^{\rm cut}$), the LP cross section is governed by a factorization formula in impact parameter ($b$) space, which captures the azimuthal decorrelation between the identified hadron and the recoiling jet in the back-to-back limit. 
A central feature of our framework is the WTA recombination scheme. 
Because the WTA axis is recoil-free, it remains insensitive to soft radiation, providing a stable reference for the hard scattering direction.
This property eliminates non-global logarithms \cite{Dasgupta:2001sh, Neill:2016vbi}, substantially simplifying the structure of the factorization formula \cite{Gutierrez-Reyes:2018qez, Chien:2020hzh}.
Building upon the soft-collinear effective theory (SCET) framework~\cite{Bauer:2000yr, Bauer:2001ct, Bauer:2001yt, Bauer:2002nz, Beneke:2002ph} and following recent developments in Refs.~\cite{Chien:2020hzh, Chien:2022wiq, Fang:2023thw, Fang:2024auf, Fu:2024fgj}, we obtain the factorized cross section
  \begin{align} \label{eq:fact_phi_jet}
      &\frac{\d\sigma_{\text{LP}}}{\d x \, \d y \, \d z \, \d^2 \vec P_{hT}  \, \d p_{\rm out}} = \int \frac{\d b}{2\pi}e^{i p_{\rm out} b/{\zeta}} \\
  &\hspace{1cm}\times   \sum_{ijk} \int \d {\xi}\,   H_{ei\to ejk}(Q,{\xi},{\zeta})\nn\\
  &\hspace{1cm}\times  {\cal B}_{i/p}({\xi},b)\,
      {\cal D}_{h/j}({\zeta},b)\,{\cal J}_{k}(b)  \, S_{ij k}(b)\,, \nn
\end{align}
where $p_{\rm out}$ is the hadron's transverse momentum relative to the scattering plane. 
The standard kinematic variables are defined as
\begin{align}
Q^2=-q^2\,,~ x=\frac{Q^2}{2P\cdot q}\,, ~ y=\frac{P\cdot q}{P\cdot l}\,,~z=\frac{P\cdot P_h}{P\cdot q}\,, 
\end{align}
with virtual photon momentum $q=l-l'$. 
The factorized ingredients $H$, $\cal B$, $\cal D$, $S$, and $\mathcal{J}$ represent the hard, TMD beam, fragmentation, soft, and WTA jet functions, respectively, where dependence on the renormalization and rapidity scales $(\mu,\nu)$ is implicit. ${\xi}$ and ${\zeta}=z+xP_{hT}^2/[z ({\xi}-x) Q^2]$ are the momentum fractions for the beam and fragmentation functions. 
Note that the dependence on the rapidity scale $\nu$ cancels in the full results of the fixed-order (FO) expansion, and we have set the factorization scale equal to the renormalization scale.
Finally, the sum $\sum_{ijk}$ runs over all contributing partonic channels, specifically photon-gluon fusion ($\gamma^* g\to q \bar q$) and QCD Compton scattering ($\gamma^* q\to q g$).
Achieving NNLO accuracy requires two-loop corrections for each factorized ingredient. 
The hard function $H_{ei\to ejk}$ encodes the tree-level and virtual corrections to the partonic process $ei \to ejk$.
We extract them from~\cite{Garland:2002ak, Gehrmann:2009vu} and find full agreement with those used in Refs.~\cite{Gehrmann:2022vuk, Gehrmann:2023zpz, NNLOJET:2025rno}.
Crucially, linearly polarized contributions from both incoming and outgoing gluons are essential for a complete description of the cross section beyond LO \cite{Catani:2010pd, Chien:2020hzh,  Gao:2023ivm}.
We employ standard TMD beam ($\mathcal{B}_{i/p}$) and fragmentation ($\mathcal{D}_{h/j}$) functions~\cite{Collins:2011zzd, Boussarie:2023izj}, applying their well-established NNLO matching onto collinear distribution functions $f_{a/p}$ and $d_{h/b}$ \cite{Catani:2011kr, Catani:2012qa, Gehrmann:2012ze, Gehrmann:2014yya, Luebbert:2016itl, Echevarria:2016scs, Luo:2019hmp, Luo:2019bmw}
\begin{align}
  {\cal B}_{i/p}\left({\xi},b\right) &=\sum_a \int_{{\xi}}^1 \frac{\d {\hat{\xi}}}{\hat{\xi}} \, {\cal I}_{i / a}(\hat{\xi}, b ) \,  f_{a/p}\left(\frac{{\xi}}{\hat{\xi}}, \mu\right)\,,   \nonumber \\
  {\cal D}_{h/j}\left({\zeta},b\right) &=
\frac{1}{{\zeta}^2}\sum_b \int_{{\zeta}}^1 \frac{\d {\hat{\zeta} }}{\hat{\zeta}} \, {\cal C}_{b / j}(\hat{\zeta}, b ) \, d_{h/b}\left(\frac{{\zeta}}{\hat{\zeta}}, \mu\right) \,, 
\end{align}
with ${\cal I}_{i / a}=\delta_{ia}\delta(1-\hat{\xi})+\mathcal{O}(\alpha_s)$ and ${\cal C}_{b / j}=\delta_{bj}\delta(1-\hat{\zeta})+\mathcal{O}(\alpha_s)$. 
The soft function $S_{ijk}$ follows the universal NNLO TMD soft function definition \cite{Echevarria:2015byo, Lubbert:2016rku}, retaining a simple analytic form because the hard scattering process is in a planar configuration  \cite{Gao:2019ojf, Chien:2020hzh, Chien:2022wiq, Gao:2023ivm, Fu:2024fgj}.
Furthermore, we adopt the $k_T$ algorithm and the WTA recombination scheme for the TMD jet functions $\mathcal{J}_k$, to be consistent with the numerical evaluation in the resolved region. 
The NNLO quark jet functions are readily available from recent calculations \cite{Gutierrez-Reyes:2019vbx, Fang:2024auf, Bell:2021dpb, Brune:2022cgr, Buonocore:2025ysd}. 
While the NNLO WTA gluon jet functions, comprising unpolarized and linearly polarized components, are currently unknown, renormalization invariance uniquely determines the two-loop logarithmic terms.
We approximate the remaining two constants of the gluon jet functions with rescaled energy-energy correlation jet functions \cite{Luo:2019hmp,Luo:2019bmw}.
Details on components of the gluon jet functions and a validation on our default choices are also given (see Supplemental Material~\cite{SM}).

\begin{figure}[t!]
    \includegraphics[width=0.9\linewidth]{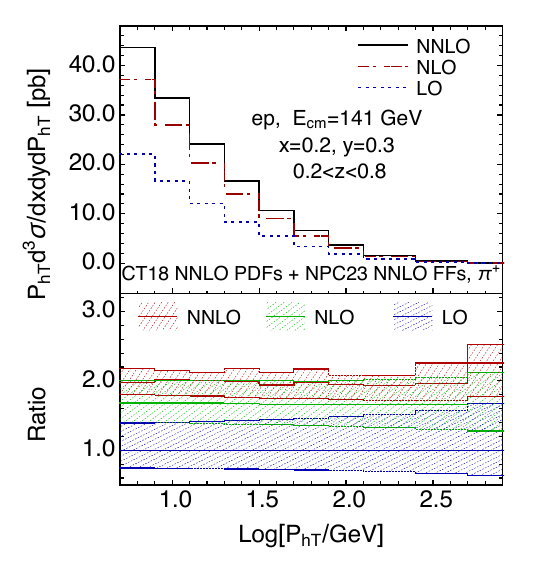}
    \caption{%
    Differential cross section as a function of the hadron transverse momentum $P_{hT}$ for charged pion production at the EIC at various orders in QCD together with scale variations, where we set $\delta\phi^{\rm cut}=0.01$.
    }
    \label{fig:pt_dist}
\end{figure}

\paragraph*{Numerical results.---}
To validate our calculation framework, we explicitly verify that the physical cross section remains independent of the arbitrary cutoff $\delta\phi^{\rm cut}$ in the limit $\delta\phi^{\rm cut}\to 0$. 
Fig.~\ref{fig:cutoff_dep} displays the numerical stability of the $\mathcal{O}(\alpha_s^3)$ (NNLO) and $\mathcal{O}(\alpha_s^2)$ (NLO) corrections to the differential cross section of charged pion production at the EIC.
The kinematics are fixed at $\sqrt{s}=141$ GeV, $x=0.2$, $y=0.3$, with integration over $0.2<z<0.8$ and $P_{hT} > z Q/10$. 
The results are shown for six representative partonic channels, including $uu$, $ug$, $gu$, $gg$, $u\bar u$, and $u d$, labeled according to the flavors of the incoming and fragmenting partons.
Among these, the $uu$ channel is always dominant due to both favored PDFs and FFs. 
In all our calculations we use the CT18 NNLO PDFs~\cite{Hou:2019efy} and the NPC23 NNLO FFs of pions~\cite{Gao:2025hlm} through all orders, with a nominal choice of $\mu=Q$ for both the renormalization and factorization scales.
The only exception is for calculations of unidentified charged hadrons, where we use the NPC23 NLO FFs~\cite{Gao:2024nkz, Gao:2024dbv} through all orders since the NNLO FFs are not available.
Note that each of the data points in Fig.~\ref{fig:cutoff_dep} is from independent Monte Carlo (MC) integrations and the shown MC uncertainties are uncorrelated.
The NLO results (red points) exhibit excellent flatness over the range $10^{-3} < \delta\phi^{\rm cut} < 10^{-1}$, confirming a precise cancellation of IR divergences, and converge to the reference values calculated with {\tt FMNLO}~\cite{Liu:2023fsq}. 
At NNLO (blue points), the results similarly converge to a constant plateau, despite the larger MC integration uncertainties from the resolved cross section.
The solid lines represent a fit to the genuine NNLO corrections together with residual power corrections~\cite{Grazzini:2017mhc, Ebert:2019zkb}.
From the fit, we estimate a systematic uncertainty (discrepancy \textit{w.r.t.} the genuine result) of less than $5\%$ ($2\%$) when a finite value of $\delta\phi^{\rm cut}=0.04\,(0.01)$ is used in the NNLO calculations for all partonic channels (see Supplemental Material~\cite{SM} for details). 
The stability observed across both quark- and gluon-initiated channels establishes the robustness of our method for high-precision phenomenology. 
In addition, we have checked that, when increasing the above unpolarized (polarized) constants by a factor of three, the NNLO corrections to the $uu$ channel change by less than $1.5\%$ ($0.02\%$) of the correction itself, which is negligible. 
\begin{figure}[t!]
    \includegraphics[width=0.9\linewidth]{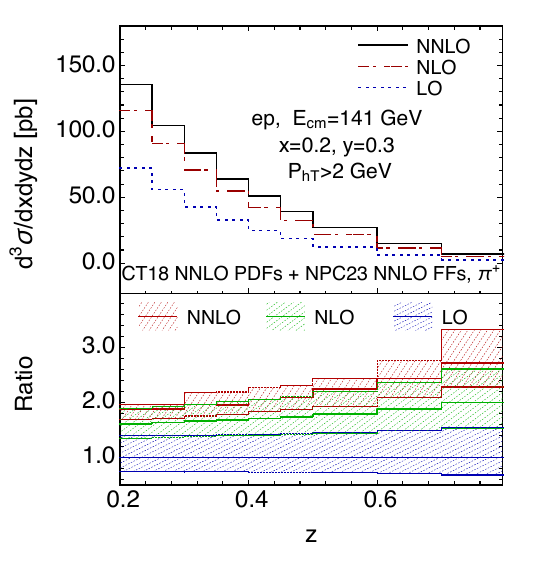}
    \caption{%
    Differential cross section as a function of the hadron momentum fraction $z$ for charged pion production at the EIC at various orders in QCD together with scale variations, where we set $\delta\phi^{\rm cut}=0.01$.
    Kinematic cuts, including a lower limit of 2~GeV on the hadron transverse momentum, are applied. 
    }
    \label{fig:z_dist}
\end{figure}
Figs.~\ref{fig:z_dist} and~\ref{fig:pt_dist}  present our NNLO predictions for the charged pion differential cross sections, tailored to the same kinematics of the future EIC used in Fig.~\ref{fig:cutoff_dep}, except for replacing the transverse momentum cut with $P_{hT} > 2$~GeV.
We set $\delta\phi^{\rm cut}=0.01$ in the NNLO calculations.
Fig.~\ref{fig:z_dist} shows the differential cross section as a function of the momentum fraction $z$. 
The upper panel displays the perturbative progression from LO (blue dotted) to NLO (red dot-dashed) and NNLO (black solid).
We observe a substantial positive correction at NNLO across the entire $z$ range, highlighting the importance of the $\mathcal{O}(\alpha_s^3)$ contributions.
The lower panel, which normalizes results to the central LO prediction, reveals the improved convergence of the perturbative expansion. 
The scale uncertainty bands (hatched regions), calculated by varying $\mu$ by a factor of two from the nominal value $Q$, undergo a significant reduction when moving from NLO to NNLO for moderate $z$ values. 
Note that the uncertainty bands represent only perturbative scale variations. 
We have checked that the systematic uncertainties from the extrapolation $\delta \phi^\text{cut} \to 0$ are subdominant compared to the residual scale variations across the shown kinematic ranges, and thus are not included in the bands.
Furthermore, the $K$-factor exhibits a distinct kinematic dependence, rising noticeably at large $z$. 
This behavior indicates that in the threshold region the soft-gluon dynamics begin to dominate~\cite{Abele:2022wuy, Goyal:2025bzf}.
Higher-order corrections beyond $\mathcal{O}(\alpha_s^3)$, for instance from soft-gluon resummation, can further reduce the perturbative uncertainties at large $z$.
Fig.~\ref{fig:pt_dist} displays the differential cross section as a function of the hadron transverse momentum $P_{hT}$, which is of particular significance as it directly probes the transition from the non-perturbative TMD regime to the perturbative collinear domain. 
The upper panel reveals a clear hierarchy in the perturbative series, with the NNLO prediction (black solid line) providing a refined description of the spectrum. 
The ratio plot in the lower panel demonstrates improved convergence of the expansion in general. 
While the jump from LO to NLO is substantial—reflecting the opening of new partonic channels—the correction from NLO to NNLO is moderate and stable. 
Crucially, the scale uncertainty (hatched bands) is largely reduced at NNLO compared to the lower orders. 
This improved theoretical precision is essential for reliably identifying the matching window between FO calculations and TMD resummation, thereby facilitating a robust extraction of the nucleon's three-dimensional structure at the EIC.
\begin{figure}[t!]
    \includegraphics[width=0.9\linewidth]{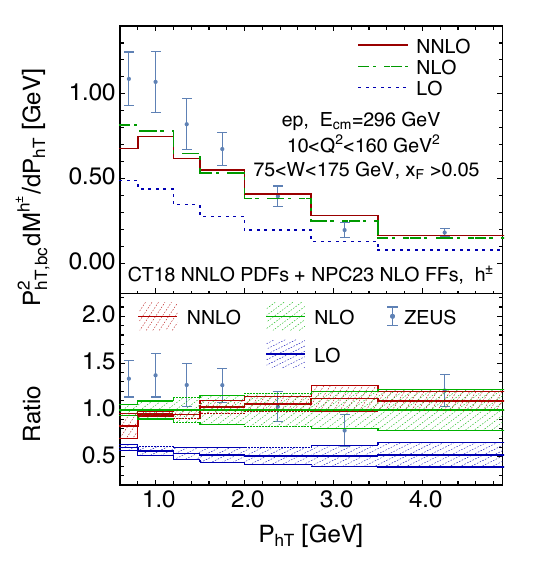}
    \caption{%
    Multiplicity distribution (multiplied by the square of the bin center) as a function of the hadron transverse momentum $P_{hT}$ for unidentified charged hadron production at HERA at various orders in QCD together with scale variations, compared to the ZEUS data~\cite{ZEUS:1995acw}, where we set $\delta\phi^{\rm cut}=0.04$ to ensure adequate control over MC uncertainties. 
    }
    \label{fig:hera}
\end{figure}

In Fig.~\ref{fig:hera}, we compare our predictions with high-precision ZEUS data \cite{ZEUS:1995acw}, collected at an $ep$ center-of-mass energy of $E_{\rm cm}=296$~GeV.
We set $\delta\phi^{\rm cut}=0.04$ in the NNLO calculations.
The plot displays the unidentified charged hadron multiplicity as a function of transverse momentum $P_{hT}$ in the kinematic region with $10< Q^2 <160$ GeV$^2$, $75< W <175$ GeV, and $x_F>0.05$, where the Feynman $x$ variable is defined as $x_F\equiv 2P_{hL}^*/W$.
The longitudinal momentum $P_{hL}^*$ is the projection of the hadron momentum onto the virtual photon direction in the center-of-mass frame of $\gamma^*p$, and $W$ is the invariant mass of the hadronic final state.  
The LO result (blue dotted) grossly underestimates the data, accounting for only $\sim 50\%$ of the observed yield. This large discrepancy highlights the limitation of LO kinematics in generating high-$P_{hT}$ recoil. 
While NLO predictions (green) significantly reduce this gap, they continue to systematically undershoot the experimental central values and show scale variations as large as the experimental errors. 
The inclusion of $\mathcal{O}(\alpha_s^3)$ corrections leads to a moderate enhancement of the normalization and reduced scale variations at large transverse momentum.
On the other hand, it is clear that for $P_{hT}\sim 1$~GeV or smaller the FO calculations are insufficient and TMD resummation~\cite{Collins:1984kg, Gao:2026tnd} is required to describe the data.

\paragraph*{Summary and Outlook.---}
In this Letter, we have presented the first NNLO QCD calculation for single inclusive hadron production in DIS at finite transverse momentum. 
By implementing the $q_T$-subtraction method with a recoil-free WTA axis, we have successfully overcome the technical bottleneck of consistently treating IR divergences in the presence of both fragmenting hadrons and jets.
Our numerical results demonstrate that NNLO corrections are substantial and significantly reduce the scale uncertainties compared to NLO. 
While residual higher-order effects remain relevant in certain extreme kinematic regions, this work provides a precision theoretical input for the hard scattering cross sections, enabling future global analyses to precisely delineate the transition region between collinear and TMD factorization and to extract non-perturbative PDFs and FFs with higher precision.
Future extensions of this work to polarized scattering will further unlock the potential of the EIC to probe the spin-momentum correlations of partons within hadrons. 
Finally, our results also serve as a fundamental step toward fully differential N$^3$LO predictions for SIDIS~\cite{Gao:2026tnd}.

\paragraph*{Acknowledgments.---}
We thank Guido Bell, Kevin Brune, Xuan Chen, Thomas Gehrmann, Ian Moult, and Hua Xing Zhu for helpful discussions.
The work of L.~D. and J.~G. is supported by the National Natural Science Foundation of China (NSFC) under Grant No.~12275173 and the Shanghai Municipal Education Commission under Grant No.~2024AIZD007.
S.~F. and D.~Y.~S. are supported by the National Natural Science Foundation of China under Grant No.~12275052, No.~12147101, No.~12547102, and the Innovation Program for Quantum Science and Technology under grant No. 2024ZD0300101.
H.~T.~L. is supported by the National Natural Science Foundation of China under Grant Nos. 12275156 and 12321005, and by the Shandong Provincial Department of Science and Technology under Project No. TQ012025001.
The computations in this paper were run on the Siyuan-1 cluster supported by the Center for High Performance Computing at Shanghai Jiao Tong University.

\bibliographystyle{apsrev4-2}
\bibliography{jet.bib}

\clearpage
\appendix
\onecolumngrid
\setcounter{equation}{0}
\renewcommand{\theequation}{S-\arabic{equation}}
 \allowdisplaybreaks
\section{Supplemental Material}

\subsection{WTA jet functions}

The subtracted WTA jet functions are defined as
\begin{align}
     J_{q}( b) &\equiv {\cal J}_{q}(b) \sqrt{S_q( b)}\,,  \\
    J_{g_c}( b) &\equiv {\cal J}_{g_c}(b) \sqrt{S_g( b)}\,. 
\end{align}
Here, the index $c$ denotes the gluon polarization. Explicitly, $c=U$ and $c=L$ correspond to the unpolarized and linearly-polarized gluon components, respectively. 
The functions ${\cal J}_{q}$ and ${\cal J}_{g_c}$ are the unsubtracted WTA quark and gluon jet functions introduced in Eq.~\eqref{eq:fact_phi_jet} of the main text, while $S_q$ and $S_g$ are the standard TMD quark and gluon soft functions. 
At NNLO, the subtracted WTA jet functions take the form
\begin{align}\label{eq: form}
     J_q &= 1 +\frac{\alpha_s C_F}{4\pi}\left[-L_b^2 + L_b(2L_\omega+3)  -\frac{5\pi^2}{6}+7-6\,\ln2 \right]\nn\\
 &+\left(\frac{\alpha_s}{4\pi}\right)^2  \Bigg[
 C_F^2\Bigg(
 \frac{L_b^4}{2}
 -L_b^3(2L_\omega+3)
 +L_b^2\left(2L_\omega^2+6L_\omega-\frac{5}{2}+6\ln 2+\frac{5\pi^2}{6}\right)\nn\\
 &+L_b\left(L_\omega\left(14-12\ln 2-\frac{5\pi^2}{3}\right)+\frac{45}{2}
 -18\ln 2-\frac{9\pi^2}{2}+24\zeta_3\right)\Bigg)\nn\\
 &+C_F C_A\Bigg(
 -\frac{22}{9}L_b^3
 +L_b^2\left(\frac{11}{3}L_\omega-\frac{35}{18}+\frac{\pi^2}{3}\right)+L_\omega\left(\frac{404}{27}-14\zeta_3\right)\nn\\
 &+L_b\left(L_\omega\left(\frac{134}{9}-\frac{2\pi^2}{3}\right)+\frac{57}{2}
 -22\ln 2-\frac{11\pi^2}{9}-12\zeta_3\right)\Bigg)\nn\\
 &+C_F T_F n_f\Bigg(
 \frac{8}{9}L_b^3
 +L_b^2\left(-\frac{4}{3}L_\omega+\frac{2}{9}\right)
 +L_b\left(-\frac{40}{9}L_\omega-10+8\ln 2+\frac{4\pi^2}{9}\right)
 -\frac{112}{27}L_\omega \Bigg)
+j_{q,\rm WTA}^{(2)} \Bigg]   \,, \\
    J_{g_U} &= 1 
     +\frac{\alpha_s }{4\pi}
 \Bigg[C_A \Bigg(
\!\!- L_b^2
+ L_b \left(2 L_\omega+\frac{11}{3}  \right)
+\frac{131}{18}- \frac{5\pi^2}{6}- \frac{22 }{3}\ln 2\Bigg)
+ T_F n_f\ \Bigg(
\!\!- \frac{4 }{3}L_b
-\frac{17}{9}+ \frac{8 }{3}\ln 2
\Bigg)\Bigg]
\nn\\
 &+\left(\frac{\alpha_s}{4\pi}\right)^2  \Bigg[
 C_A^2 \Bigg(
\frac{L_b^4}{2}
+ L_b^3 \left(- 2 L_\omega-\frac{55}{9}\right)
+ L_b^2 \left(
 2 L_\omega^2+ 11 L_\omega
-\frac{23}{18}+ \frac{7\pi^2}{6}+ \frac{22 }{3}\ln 2
\right)\nn\\
&+ L_b \left(
 L_\omega \left(
\frac{265}{9}- \frac{7\pi^2}{3}- \frac{44 }{3}\ln 2\right)
+\frac{1729}{27}
- \frac{121\pi^2}{18}
- \frac{484 }{9}\ln 2+ 12 \zeta_3
\right)
+ L_\omega \left(
\frac{404}{27}- 14 \zeta_3
\right)\Bigg)
\nn\\
&+ C_A T_F n_f\Bigg(
\frac{20 }{9}L_b^3
+ L_b^2 \left(
- 4 L_\omega
-\frac{17}{3}
- \frac{8 }{3}\ln 2
\right)
\nn\\
&+ L_b \left(
 L_\omega \left(
-\frac{74}{9}+ \frac{16 \ln 2}{3}\right)
-\frac{1042}{27}+ \frac{22\pi^2}{9}+ \frac{352 }{9}\ln 2
\right)
- \frac{112 }{27}L_\omega
\Bigg)
\nn\\
&- 4 C_F T_F n_f L_b
+ T_F^2 n_f^2  \Bigg(
\frac{16 }{9}L_b^2
+ L_b \left(
\frac{136}{27}- \frac{64 }{9}\ln 2
\right)
\Bigg)
+j_{g_U,\rm WTA}^{(2)}
\Bigg]\,, \label{eq:JgU}
\\ 
    J_{g_L} &= 
     \frac{\alpha_s }{4\pi}
     \left[
\frac{1}{3}C_A
- \frac{2}{3} T_F n_f
\right] \label{eq:JgL}
\\
&+ \left(\frac{\alpha_s}{4\pi}\right)^2 \Bigg[
 C_A^2 \left(
\!\!-\frac{L_b^2}{3}
+ L_b \left(
\frac{2 }{3}L_\omega+\frac{22}{9} 
\right)
\right)
+ C_A T_F n_f\left(
\frac{2 }{3}L_b^2
+ L_b \left(
-\frac{52}{9}
- \frac{4 }{3}L_\omega
\right)
\right)
+ \frac{16}{9} T_F^2 n_f^2   L_b
+j_{g_L,\rm WTA}^{(2)}
\Bigg]\,, \nn  
\end{align}
respectively, with $j_{q,\rm WTA}^{(2)}\approx 6.60$~\cite{Fang:2024auf}, $L_b=\ln(\mu^2 b^2 e^{2\gamma_E}/4)$ and $L_\omega=\ln(\mu^2/\omega^2)$. Here $\omega$ represents twice the jet energy.
As for the two constants $j_{g_U,\rm WTA}^{(2)}$ and $j_{g_L,\rm WTA}^{(2)}$ in Eqs.~\eqref{eq:JgU} and \eqref{eq:JgL}, to our knowledge, their exact values are currently unknown.
We therefore approximate them by rescaling the EEC gluon jet function constants~\cite{Luo:2019bmw} by the ratio of the constant terms in the WTA and EEC quark jet functions~\cite{Luo:2019hmp}. Note that the anomalous dimensions should be the same between EEC and WTA jet functions based on the factorization formula. In other words,
\begin{align}
    \label{eq: appr_jg2}
    j_{g_U, \text{WTA}}^{(2)} &= j_{g_U, \text{EEC}}^{(2)} \times \frac{j_{q, \text{WTA}}^{(2)}}{j_{q, \text{EEC}}^{(2)}}\approx-12.9\,, \nn\\
    j_{g_L, \text{WTA}}^{(2)} &= j_{g_L, \text{EEC}}^{(2)} \times \frac{j_{q, \text{WTA}}^{(2)}}{j_{q, \text{EEC}}^{(2)}}\approx0.97\,. 
\end{align}
To assess the reliability of this approximation, we perform the same test at NLO, i.e.,
\begin{align}
    j_{g_U, \text{WTA}}^{(1)} &= j_{g_U, \text{EEC}}^{(1)} \times \frac{j_{q, \text{WTA}}^{(1)}}{j_{q, \text{EEC}}^{(1)}}\approx-17.42\,, \nn\\
    j_{g_L, \text{WTA}}^{(1)} &= j_{g_L, \text{EEC}}^{(1)} \times \frac{j_{q, \text{WTA}}^{(1)}}{j_{q, \text{EEC}}^{(1)}}\approx-0.33\,, 
\end{align}
to be compared with the exact results
\begin{align}
    j_{g_U, \text{WTA}}^{(1)}&=C_A\left(\frac{131}{18} - \frac{5\pi^2}{6} - \frac{22}{3}\ln{2}\right)+  T_F n_f \left(-\frac{17}{9} + \frac{8}{3}\ln{2}\right) \approx -18.19\,, \nn\\
   j_{g_L, \text{WTA}}^{(1)}&= \frac{1}{3}C_A-\frac{2}{3}T_F n_f  \approx -0.67 \,.
\end{align}
Here, $C_A=3$, $T_F=1/2$ and $n_f=5$. At NLO, this approximation reproduces the correct order of magnitude, although it underestimates the exact value by roughly a factor of two for the linearly polarized gluon.
We have checked that increasing the unpolarized (polarized) value in Eq.~(\ref{eq: appr_jg2}) by a factor of three changes the NNLO correction in the $uu$ channel shown in Fig.~\ref{fig:cutoff_dep} of the main text by less than $1.5\%$ ($0.02\%$) of the correction itself. 

\begin{figure}[t!]
    \includegraphics[width=0.5\linewidth]{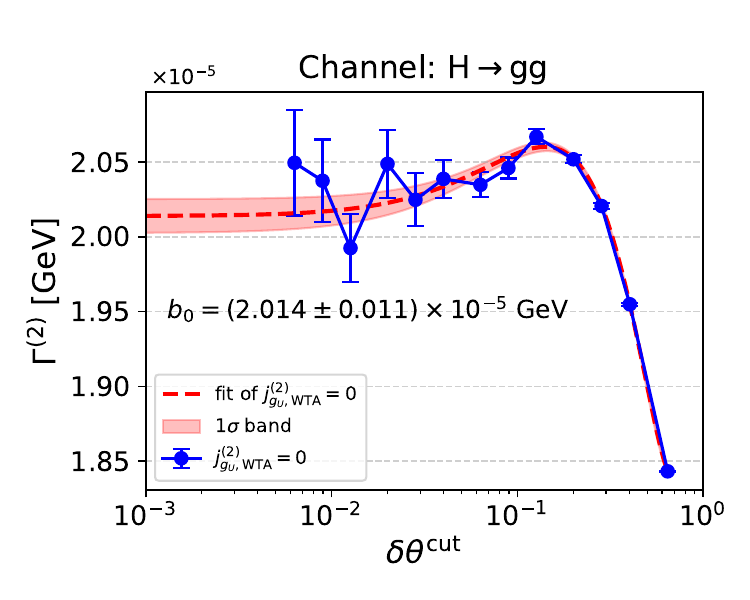}
    \caption{%
    NNLO corrections (blue points) to the partial width for the Higgs boson decay into two gluons as a function of the slicing parameter $\delta\theta^{\rm cut}$. The dashed red curve and the associated $1\sigma$ error band show the fit to the asymptotic limit $\delta\theta^{\rm cut}\to 0$. The parameter $b_0$ denotes the extrapolated value of $\Gamma^{(2)}$ in this limit.
    }
    \label{fig:jgu2}
\end{figure}

On the other hand, the value of $j_{g_U, \text{WTA}}^{(2)}$ can be extracted from the decay of the Higgs boson into two gluons.
When applying a similar $q_T$ subtraction method to the calculation of the partial width, the NNLO correction $\Gamma^{(2)}(H \to gg)$ depends linearly on $j_{g_U, \text{WTA}}^{(2)}$, and the corresponding coefficient can be calculated directly.
To determine the value of $j_{g_U, \text{WTA}}^{(2)}$, as shown in Fig. \ref{fig:jgu2}, we perform NNLO calculations with vanishing $j_{g_U, \text{WTA}}^{(2)}$ and analyze the convergence with respect to the slicing parameter $\delta\theta^{\rm cut}$.
The numerical results (blue points) display moderate power corrections at larger $\delta\theta^{\rm cut}$.
We fit these data points using the ansatz~\cite{Grazzini:2017mhc, Ebert:2019zkb}
\begin{align}
    \Gamma^{(2)}(\delta\theta^{\rm cut})=(\delta\theta^{\rm cut})^2(a_3\ln^3\delta\theta^{\rm cut}+a_2\ln^2\delta\theta^{\rm cut}+a_1\ln\delta\theta^{\rm cut}+a_0)+b_0\,,
\end{align}
where $b_0$ is the extrapolated value of $\Gamma^{(2)}$ in the limit $\delta\theta^{\rm cut} \to 0$, and $a_i$ are free parameters fitted together with $b_0$.
By comparing $b_0$ with the known analytical NNLO partial width~\cite{Baikov:2006ch}, we extract $j_{g_U, \text{WTA}}^{(2)}=-9.4 \pm 3.1$, which is consistent with the approximation in Eq.~(\ref{eq: appr_jg2}). 
As indicated by the uncertainty of the fitted $b_0$ (also shown as error bars in Fig.~\ref{fig:jgu2}), this numerical extraction precision is sufficient for our current phenomenological study, since this uncertainty propagates to an impact of less than $1\%$ on the NNLO corrections.

We now briefly review the factorization formula for the transverse momentum decorrelation in the $H \to gg$ process, and outline the fixed-order expansion of the decay width. In $b$-space, the factorization theorem reads
\begin{align}\label{eq:epem_res}
    \frac{\d\Gamma}{ \d q_{T}}=
      \Gamma_0\, H_{g}(m_H^2)\, C_t^2(m_t^2)\,  q_T\!\! \int_0^\infty  b_T \, \d b_T  J_0( q_{T} b_T)  \left[ { J}_{g_U}^2( b_T)+{ J}_{g_L}^2( b_T)\right] \,, 
\end{align}
where $J_0$ is the Bessel function of the first kind. The leading-order decay width is
\begin{align}
    \Gamma_0 = \frac{  G_F \alpha_s^2 m_H^3}{36\sqrt{2}\pi^3}\,.
\end{align}
It is instructive to perform the fixed-order expansion of the resummed result in Eq.~\eqref{eq:epem_res}. To this end, we define the integrated decay width below the cutoff $\delta\theta^{\rm cut}$ as
\begin{align}\label{eq:epem_intres}
    \Gamma_{\rm LP}(\delta\theta^{\rm cut}) \equiv \int_0^{m_H \sin\frac{\delta\theta^{\rm cut}}{2}} \d q_T \, \frac{\d \Gamma}{\d q_T}\,. 
\end{align}
Here the perturbative expansion of $\Gamma_{\rm LP}$ at NNLO can be written as
\begin{align}
\frac{\Gamma_{\rm LP}(\delta\theta^{\rm cut})}{\Gamma_0} &=    1
+      \frac{\alpha_s }{4\pi} \Bigg[
 C_A \Bigg(\!\!
-2L_{\theta}^2
+L_{\theta}\left(-4L_H+\frac{22}{3}\right)
-2L_H^2
+\frac{221}{9}
+\frac{2\pi^2}{3}
-\frac{44}{3}\ln2
\Bigg)
\\
&+ T_F n_f \Bigg(\!\!
-\frac{8}{3}L_{\theta}
-\frac{34}{9}
+\frac{16}{3}\ln2
\Bigg)
-6 C_F\Bigg]
\nn\\
&+ \left(\frac{\alpha_s}{4\pi}\right)^2 \Bigg[
 C_A^2 \Bigg(
2L_{\theta}^4
+L_{\theta}^3\left(
8L_H-\frac{176}{9}
\right)
+L_{\theta}^2\left(
12L_H^2
-\frac{110}{3}L_H
-\frac{71}{3}
-\frac{2\pi^2}{3}
+\frac{88}{3}\ln2
\right)
\nn\\
&+L_{\theta}\Bigg(
8L_H^3
-\frac{44}{3}L_H^2
+L_H\left(
-128
-\frac{4\pi^2}{3}
+\frac{176}{3}\ln2
\right)
+\frac{2293}{9}
-\frac{22\pi^2}{9}
-\frac{484}{3}\ln2
-8\zeta_3
\Bigg)\nn\\
&+2L_H^4
+\frac{22}{9}L_H^3
+L_H^2\left(
-64
-\frac{2\pi^2}{3}
+\frac{88}{3}\ln2
\right)
+L_H\left(
-24
-\frac{88\pi^2}{9}
-8\zeta_3
\right)
+14L_t
\nn\\
&+\frac{37377}{108}
+\frac{424\pi^2}{9}
-\frac{41\pi^4}{36}
-\frac{6842}{27}\ln2
-22\pi^2\ln2
+\frac{484}{9}\ln^2 2
+\frac{418}{9}\zeta_3
\Bigg)
\nn\\
&+ C_A C_F \Bigg(
12L_{\theta}^2
+L_{\theta}\left(
24L_H-44
\right)
+12L_H^2
-22L_t
-184
-4\pi^2
+88\ln2
\Bigg)
\nn\\
&+ C_A T_F n_f\Bigg(
\frac{64}{9}L_{\theta}^3
+L_{\theta}^2\left(
\frac{40}{3}L_H
-\frac{52}{3}
-\frac{32}{3}\ln2
\right)
+L_{\theta}\Bigg(
\frac{16}{3}L_H^2
+L_H\!\left(
24-\frac{64}{3}\ln2
\right)
\nn\\
&-\frac{1234}{9}
+\frac{8\pi^2}{9}
+\frac{352}{3}\ln2
\Bigg)
-\frac{8}{9}L_H^3
+L_H^2\left(
12-\frac{32}{3}\ln2
\right)
+L_H\left(
16+\frac{32\pi^2}{9}
\right)
-\frac{9833}{81}
-\frac{101\pi^2}{9}
\nn\\
&+\frac{3236}{27}\ln2
+8\pi^2\ln2
-\frac{352}{9}\ln^2 2
-\frac{440}{9}\zeta_3
\Bigg)
+ C_F T_F n_f \Bigg(
8L_{\theta}
+8L_H
+16L_t
-32
-32\ln2
+32\zeta_3
\Bigg)
\nn\\
&+ T_F^2 n_f^2 \Bigg(
\frac{16}{3}L_{\theta}^2
+L_{\theta}\left(
\frac{136}{9}
-\frac{64}{3}\ln2
\right)
+\frac{325}{81}
-\frac{272}{27}\ln2
+\frac{64}{9}\ln^2 2
\Bigg)
+36C_F^2
-\frac{8}{3}C_F T_F
-\frac{5}{3}C_A T_F
+2j_{g_U}^{(2)}
\Bigg]\,,\nn 
\end{align}
with $L_H=\ln (m_H^2/\mu^2)$, $L_{\theta}=-2\ln[\sin(\delta\theta^{\rm cut}/2)]-L_H$, and $L_t=\ln(\mu^2/m_t^2)$.
Combining the contribution below and above the cutoff, we have  
\begin{align}
    \Gamma(\delta\theta^{\rm cut}) =\Gamma_{\rm LP}(\delta\theta^{\rm cut}) +\int_{\delta \theta^{\rm cut}}^{\delta \theta^{\rm max}} \d \delta\theta \frac{\d\Gamma}{ \d\delta \theta}\,.
\end{align}

\subsection{Power corrections and finite-cutoff error}

\begin{table}[t!]
    \centering
    \renewcommand{\arraystretch}{1.6}
    \begin{tabular}{|c|c|c|c|c|c|c|}
        \hline
        \diagbox{$\delta\phi^\text{cut}$}{Channel} & $gg$ & $gu$ & $ud$ & $ug$ & $uu$ & $u\bar{u}$ \\
        \hline
        $0.04$ & 0.55\% \ (1.98\%) & 4.04\% \ (5.60\%) & 2.39\% \ (3.53\%) & 3.89\% \ (4.21\%) & 0.37\% \ (4.72\%) & 2.28\% \ (3.21\%) \\
        \hline
        $0.02$ & 0.21\% \ (0.94\%) & 1.82\% \ (2.74\%) & 0.65\% \ (1.85\%) & 1.86\% \ (2.36\%) & 0.95\% \ (2.40\%) & 0.77\% \ (1.66\%)  \\
        \hline
        $0.01$ & 0.22\% \ (0.48\%) & 0.74\% \ (1.11\%) & 0.15\% \ (0.94\%) & 0.79\% \ (1.41\%) & 0.61\% \ (1.23\%) & 0.24\% \ (0.76\%) \\
        \hline
        $0.004$ & 0.09\% \ (0.29\%) & 0.21\% \ (0.56\%) & 0.01\% \ (0.40\%) & 0.23\% \ (1.05\%) & 0.23\% \ (1.19\%) & 0.05\% \ (0.17\%) \\
        \hline
    \end{tabular}
    \caption{Estimated finite-cutoff uncertainties of the NNLO cross section obtained from fits to the power corrections for several choices of $\delta\phi^\text{cut}$. The numbers in parentheses show the corresponding NLO deviations from the exact results.}
    \label{tab:bias}
\end{table}

The solid cyan curves in Fig.~\ref{fig:cutoff_dep} of the main text represent fits that include power corrections, which take the following form~\cite{Grazzini:2017mhc, Ebert:2019zkb}
\begin{align}
    \sigma^{(2)}(\delta\phi^\text{cut})=\left(\delta\phi^\text{cut}\right)^2\left(c_3\ln^3\delta\phi^\text{cut}+c_2\ln^2\delta\phi^\text{cut}\right)+d_0\,.
\end{align}
Here we use a three-parameter expression containing two leading power corrections.
Similar results are obtained when subleading terms are also
included.
Crucially, since the data points at different $\delta\phi^\text{cut}$ are generated from independent Monte Carlo integrations, their statistical uncertainties are uncorrelated. 
The fits are therefore performed via a $\chi^2$ minimization weighted by these independent statistical errors. For all six channels, we obtain $\chi^2/N_{\rm d.o.f.} \lesssim 1$. 
The fitted constant $d_0$ corresponds to the extrapolated value of the NNLO correction in the limit of $\delta\phi^\text{cut} \to 0$. 

Tab.~\ref{tab:bias} summarizes estimates of the finite-cutoff uncertainties by comparing the fitted $\sigma^{(2)}(\delta\phi^\text{cut})$ at a finite cutoff with $d_0$, for $\delta\phi^\text{cut}=0.04$, $0.02$, $0.01$ and $0.004$, and for all six partonic channels. 
The quoted percentages represent the relative differences between these two values, and hence provide an estimate of the systematic uncertainty induced by a fixed finite $\delta\phi^\text{cut}$.
For comparison, in parentheses we also give the relative differences between $\sigma^{(1)} (\delta\phi^\text{cut})$ and the exact NLO corrections for the same cutoff choices.

\end{document}